\documentclass[12pt,preprint]{aastex}

\shorttitle{Metal Production in Galaxy Clusters}
\shortauthors{Bregman, Anderson \& Dai}

\begin{document}


\title{Metal Production in Galaxy Clusters: The Non-Galactic Component}


\author{Joel N. Bregman, Michael E. Anderson}
\affil{Department of Astronomy, University of Michigan, Ann Arbor, MI  48109-1042}
\email{jbregman@umich.edu, michevan@umich.edu}

\and

\author{Xinyu Dai}
\affil{Department of Physics and Astronomy, The University of Oklahoma, Norman, OK 73019}
\email{xdai@ou.edu}



\begin{abstract}
The metallicity in galaxy clusters is expected to originate from the stars in
galaxies, with a population dominated by high mass stars likely being the most
important stellar component, especially in rich clusters.  We examine the
relationship between the metallicity and the prominence of galaxies as
measured by the star to baryon ratio, M$_*$/M$_{bary}$.  Counter to expectations, 
we rule out a metallicity that is proportional to M$_*$/M$_{bary}$, where the 
best fit has the gas phase metallicity decreasing with M$_*$/M$_{bary}$, or the
metallicity of the gas plus the stars being independent of M$_*$/M$_{bary}$.  
This implies that the population of stars responsible for the metals is largely 
proportional to the total baryonic mass of the cluster, not to the galaxy mass within the cluster.  
If generally applicable, most of the heavy elements in the universe were not 
produced within galaxies.
\end{abstract}





\section{Introduction}

While nucleosynthesis is undoubtedly responsible for the production of the
elements, there is less certainty regarding the stellar populations involved, as
well as the time and location of elemental production.  Historically, these
issues have been studied within individual galaxies, an ideal approach if
they are nearly closed box systems.  However, studies find that galaxies 
have lost the majority of their baryons
\citep{mcga05} and some undetermined amount of metals.  For example, a
galaxy like the Milky Way is missing 75\% of its original baryon content while
for M33, the deficit is about 90\% \citep{ande10}. 

Galaxy clusters provide another useful system for studying metallicity
evolution, as moderate and rich clusters retain most of their baryons (e.g., 
\citealt{bohr09}).  Baryons lost from galaxies are captured in the hot
cluster medium, which is effectively studied in X-rays (e.g., \citealt{wern08}). 
The first order of business is to assess whether the stars in galaxies have
produced the observed metallicity.  Simple estimates indicate a problem
in that there are too few stars to account for the heavy element mass 
\citep{arim97,port04,loew06}.  We estimate that the metallicity of stars in
galaxies, averaged over a standard luminosity function, is about 0.5 solar. 
Galaxies comprise about 5-10\% of the baryons in rich clusters, so if the metals
in these stars were spread through the cluster, we expect the hot gas
metallicity to be about 5\% solar, but the typical metallicity is 40\% solar (e.g., 
\citealt{snow08}), about an order of magnitude higher. 

This issue was examined by several authors \citep{port04,loew06}, who did careful
calculations of metallicity production from the various types of supernovae,
using a cosmic star formation history (review of techniques by
\citealt{borg08}).  \citet{loew06} concludes that with any of the standard
initial mass functions, the metals produced by the stars in galaxies cannot
produce the cluster metals.  He posits the existence of a population of
mainly high-mass stars to produce the cluster metallicity and he investigates the time evolution. 
\citet{port04}  reach a similar conclusion and adopt the IMF proposed by
\citet{arim97}, which is top-heavy (\citealt{fabj08} use a top-heavy IMF 
and a normal IMF).  A dissenting view is given by \citet{siva09} who find that
for some clusters, the metals might have originated in galaxies.  However,
several of their adopted values are relatively extreme compared to the work of
\citet{loew06} and \citet{port04}.

In a related paper \citet{ande10}, we argued that field spirals are missing their
baryons due to the presence of an early stellar 
population dominated by high-mass stars, which heats and enriches the gas
before most of it collapsed into proto-galaxies.  In this picture, which is driven by
simple observational constraints, galaxies are baryon-poor because the gas
never fell into galactic dark matter potential wells.  Preheating (and 
pre-enrichment) by this early population would need to occur near the onset of galaxy 
formation, which occurs before cluster formation.  
Therefore, this picture should apply to systems that eventually become galaxy clusters.

It is natural to expect that the mass of the high-mass weighted early population of stars is proportional to the mass
of a galaxy, so the collective mass of this population in a cluster should be proportional
to the total stellar mass seen in galaxies today.  Translating this into a
metallicity prediction, the metallicity of the cluster should be proportional to
the ratio of the stellar mass to the total baryon mass, M$_*$/M$_{bary}$.  
This is the same expectation if the metals were produced and expelled from
the cluster galaxies.  We show that this basic
expectation is excluded by the data, potentially leading to a profound change in our
understanding of metallicity production in clusters and in the universe.

\section{Object Selection and Data Products}

For this study, we need to measure four parameters of galaxy clusters:  the
gaseous mass; the stellar mass; the total gravitating mass; and the metallicity. 
From these quantities, we can assess the total baryon fraction of clusters,
along with the ratio of the stellar mass to the total baryonic mass, and
compare it to the metallicity.  These quantities are most easily measured for
low redshift clusters, which are the objects used.  There are a variety of
measurements of the properties of low-redshift clusters, but only a limited
number measure the mass to large radii (r$_{500}$) and are close
enough to measure the stellar mass content from sources such as the 2MASS
data set \citep{koch03,lin04a,vikh09,dai10}.  We discussed many of these
objects in a separate work \citep{dai10}, from which we select clusters that
have retained most of their baryons, so have similar total baryon fractions
(Table 1); these clusters appear to be relaxed.  The gaseous masses are
determined by fitting a functional form to the surface brightness
distribution.  The functional form begins with a $\beta$-model, which can be modified so that it steepens at large radii plus another component in the center \citep{vikh09}. Temperatures are determined from the data using spectral fits (i.e., APEC models),
and this temperature information is important in determining the total
gravitating mass of the system.  Metallicities are normalized to \citet{ande89}
where the relative abundances are held at these solar ratios.  
In determining a metallicity, the most prominent lines are from Fe, which is
the most important element statistically.  Henceforth, we use [Fe/H] as a proxy for 
metallicity, where the solar elemental ratios are assumed.
A $\Lambda$CDM concordance cosmology is assumed with $H_0$ = 73 km
s$^{-1}$ Mpc$^{-1}$, ${\Omega}_m$ = 0.26, and ${\Omega}_{\Lambda}$ = 0.74.

Metallicity has been obtained as a function of radius for many of our systems
by us and others, from which one can infer an emission-weighted metallicity or
a mass-weighted metallicity (\citealt{ande09} and references therein).  The
difference is generally not large, so we use the emission-weighted metallicity,
as this is the only quantity available for clusters that are more gas-poor. 
The primary complication in making this measurement is accounting for the 
presence of cooling cores, which can bias the emission-weighted metallicity 
upwards by introducing additional flux from the enriched core, usually
coincident with a central galaxy (e.g., \citealt {rasm09}). 
We therefore present two metallicities for each cluster, one emission-weighted 
over all of the X-ray emission, and one that excludes emission from the core 
of the cluster, as described below.

Correctly accounting for a cool core requires careful deprojection of the 
emission, but we lack the data for such an analysis in all cases, and so 
perform a simpler correction. We examined the universal temperature and 
abundance profiles of \citet{rasm07} for galaxy groups and \citet{bald07} 
for hot (T $\ge 6$ keV) galaxy clusters. Both groups found temperature 
profiles that slowly rise towards the core, then decline more sharply 
starting at $R \sim 45 T^{1/2}$ kpc, where T is the emission-weighted 
temperature of the ICM in keV. Most of the objects in our sample have 
temperatures within the range of values considered in these papers, so the 
same profiles should also apply. We therefore attempt to exclude emission 
from within a projected radius of $45 T^{1/2}$ kpc in each cluster to 
produce an emission-weighted iron abundance without the bias of a bright core (T is in keV units). 
This is straightforward for the clusters with measurements from \citet{snow08}, 
since they provide measurements of the temperature, metallicity, and flux 
in a series of annuli for each cluster, which allows us to verify that 
a cooling core exists and then to remove the inner annuli and recompute 
the flux-weighted metallicity. We perform a similar analysis on 3C442A, 
for which we have measurements of temperature and metallicity as a function 
of radius (Sun 2009, private communication), and for RGH80, using the 
single-temperature fits of \citet{xue04}. For these two systems, we estimate the expected 
flux in each annulus by assuming the ICM gas density follows a $\beta$-model 
with $\beta = 0.65$. In Abell 1275 the metallicities are from our analysis of
the Chandra data.  For A160, and A2462 \citep{jeth05}, we have only global 
measurements of the metallicity, so we estimate the effect of a cooling core. 
Based on the other clusters in our sample, the the exclusion of a bright
core reduces the iron abundance by $25\%$, so we use this correction. 

The final quantity is the stellar mass, which is obtained by identifying the
galaxies and measuring their magnitudes to a radius of r$_{200}$, which
includes nearly all of the stellar light.  The data set for the galaxies used here is
the 2MASS data base, so there is a magnitude limit to galaxy identification. 
This leads to sampling only part of the galaxy luminosity function, so a correction is
applied to account for galaxy incompleteness and Poisson bias for a simple 
halo occupation model, as previously discussed \citep{koch03,lin04a,dai07}.  
The uncertainties assigned are conservative in that they are larger than 
those given in \citet{lin04a}.
   For some of the analysis, we include the metals in the stars, so we assumed a solar metallicity for that correction as applied to the total stellar mass.

This procedure does not include a contribution from intracluster light (ICL). 
Measures of the intracluster light were determined by \citet{krick07} for 10
galaxy clusters, where their fractional ICL results are consistent with a single
value of 11.1 $\pm$ 1.3\% (${\chi}^2$ minimization).  For individual clusters, the B and
r weighted ICL values range from 4.8 $\pm$ 3.1\% for Abell 2556 to 21.3 $\pm$
6.6\% for Abell 4059 and there is no obvious correlation of the ICL percentage
with cluster properties.  This measurement agrees with the value from
theoretical modeling by \citet{rudi06} and with the fraction of host-less
supernovae in clusters, which is quoted as 20\% (+20/-12) by \citet{galy03}
and $\sim$20\% by \citet{sand08}.  Another study that includes ICL is that of
\citet{gonzo07}, but they combine the ICL and the light from the brightest
central galaxy, so this does not provide an independent measure of the ICL. 
They imply that the ICL may increase toward poorer clusters.  In the following
analysis, we discuss the effects of ICL that comprise 10-20\% of the stars, as
well as an increase of the ICL toward poorer clusters.

\section{Metallicity as a Function of Baryon Fraction}

With the set of clusters for which we have good data, we examined the
relationship between the metallicity of the gas and the fraction of the baryons in the
form of stars (Fig 1).  A proportional relationship between these quantities is the
nominal expectation, but the best fit, Z = 2.44 M$_*$/M$_{bary}$, has a
badly unacceptable ${\chi}^2$ = 729 for 10 dof, so it is ruled out.  Upon removing RGH 80, the object
with the highest value of M$_*$/M$_{bary}$, the fit is still ruled out, as
${\chi}^2$ = 413 for 9 dof.  The data are characterized by a metallicity 
that is slowly declining as a function of cluster mass, while the importance 
of the stellar mass increases toward lower cluster mass. 

We consider another model to explain the data, one where part of the metals
are associated with stars B(M$_*$/M$_{bary}$), and the other where some
fraction of the metals has an origin independent of the current visible stars,
Z$_0$.  This second component can be understood as an early population 
component (e.g., dominated by high mass stars) that is not proportional to the galaxy mass.  This is expressed
as a linear equation, Z = Z$_0$ + B$\,$(M$_*$/M$_{bary}$).
We made fits to two sets of metallicities, a single emission-weighted metallicity for
the cluster and a metallicity with the core excluded.  The linear fits in
the two cases are similar, but the fit is better for the metallicities without
the core, so we concentrate on those results (Fig. 1b, 2).  For the case without
the metal contribution from the stars, the best-fit is a negative slope, 
Z = 0.37  - 0.70(M$_*$/M$_{bary}$), with errors on Z$_0$ and B of 0.02 and 
0.23; ${\chi}^2$ = 18.5 for 10 dof, which is acceptable.  
For the case with the stellar metal contibution included, the best fit has 
a flat slope, Z = 0.37  + 0.10(M$_*$/M$_{bary}$), with errors on Z$_0$ and B of 0.02 and 
0.22; ${\chi}^2$ = 14.4 for 10 dof.
The second fit is consistent with a constant metallicity, but the fit without
the stellar metalliity is inconsistent with a constant metallicity solution.

The inclusion of ICL either does not change our result or make it more restrictive. 
When calculating the stellar component, we used a K band M/L of 0.95.  Had
we used the IMF suggested by \citet{chab03}, the M/L would be about 0.72,
resulting in a stellar mass that is 24\% lower.  On the other hand, inclusion of
ICL would increase the stellar mass by 11\%, using \citet{krick07}.  Either of
these effect lead to minor scaling changes in M$_*$/M$_{bary}$, but do not
change the results.  If the ICL increases significantly toward poorer clusters,
which preferentially occur in the high M$_*$/M$_{bary}$ part of the diagram,
the maximum allowable slope for the gas metallicity is lowered as is the 
fraction of metals than can be associated with galaxies.

This general result, that the metallicity is constant or decreases as the stellar fraction 
increases, has been known for over a decade.  An increase in the ratio of the 
stellar mass to the gas mass with decreasing cluster temperature was first
demonstrated by \citet{david90}.  A modest decrease of cluster metallicity with
decreasing cluster temperature was discussed by \citet{fuka98}.  Our work
confirm these trends but uses more uniform stellar metallicities, gas masses,
and total gravitating masses, all measured to large radii.  For our conclusions to
be invalid, a large body of work by others would need to be dramatically wrong.

Although we used systems in which the baryons are mostly retained by the 
clusters, the situation is not different if extended to galaxy groups in general.
In their recent survey of 15 groups of galaxies, \citet{rasm09} find that the 
metallicity of the gas beyond the central galaxy is lower than in clusters,
a trend that is consistent with our result (Fig. 1b).

The implications of this result are remarkable in that it suggests that not only
are most of the metals in the richest clusters produced by the an IMF dominated by 
high mass stars, but that the amount of mass in this component is not
proportional to the stellar mass in galaxies.  The need to produce most of the metals
in an IMF that is heavily weighted on the high-mass end was
previously discussed \citep{arim97,loew06,port04}.  This inference is
supported by the elemental ratios within rich clusters, which favor metal
production by an ensemble of high-mass metal-poor progenitors
\citep{baum05}.  The mass of this stellar population would appear to be
roughly proportional to the mass of gas in a cluster.  This conclusion would
explain the mean cluster metallicity, but not the metallicity gradient within a
cluster.  For that to be explained with this stellar component, the frequency of
such SNe would need to be biased toward higher density regions.  This is
similar to the arguments that were put forward for biased galaxy formation, so
this might be considered biased star formation with a high-mass IMF.

Clusters of galaxies were used in this study because they represent a "closed
box" piece of the universe.  The situation in clusters is likely to be
representative of the universe as a whole, in which case, most of the heavy
elements in the universe were produced outside of galaxies by an early
population of supernovae.  If correct, this represents a fundamental
change in our understanding for the origin of the elements.  This moderately
enriched material (0.1-0.4 solar) would fall into galaxies, producing a disk with
few low-metallicity stars (a previously proposed solution to the G-dwarf
problem).  The increase of the metallicity within galaxies is of course due to
star formation, mass loss and supernovae within the galaxy.

In addition to the above issues, some additional evidence supports our model.
\citet{ehle09} analyzed cluster metallicity profiles from z = 0.14 to z = 0.89 and
from the lack of evolution, they conclude that the metallicity distribution was
established at high redshift. \citet{david08} note that Fe is more extended
than the light of stars, which could imply that the galaxies and the metals are
not closely coupled (they suggest an alternative model where AGN heating is
responsible). \citet {somm08} find that at z $\sim$ 3, less than 20-25\% of the
oxygen is associated with galaxies, which is also consistent with most of the
metals being produced outside of galaxies.

If there was a predominantly high-mass population of stars outside galaxies,
the initial mass function of these stars could be constrained by the ICL,
provided one could separate the ICL from galaxy-galaxy interactions with that
from the high-mass stars.  The present values of the ICL must
provide a limit to the slope of that IMF, but future work may be able to
constrain it more directly.  Another constraint on the nature of the metal
production has been the variation of the elemental ratios with radius
\citep{dupke00,baum05,rasm09}.  In groups, the Si/Fe ratio increases with distance
away from the central galaxy, indicating that Type Ia are less important at
larger radii \citep{rasm09}.  There is significant room for improvement in these types of
studies, by using more elements, and especially for the hot clusters.

In obtaining this result, only a dozen clusters were used, a shortcoming that
can be rectified in future work.  Rich clusters are well-represented and the
number of such objects can be increased with straightforward archival work of
X-ray data; these tend to have low values of M$_*$/M$_{bary}$.  The greater
weakness is the number of galaxy clusters with high values of
M$_*$/M$_{bary}$.  These tend to be of lower mass, relatively gas-poor and
cooler systems, so the X-ray emission is weaker, making metallicity
determinations from X-ray data more challenging.  Additional X-ray data will
be needed to make progress here.  Finally, more accurate values of M$_*$ can
be determined with deep imaging, as typically only two dozen galaxies define
L$_*$ in a cluster.  While these are the the brightest galaxies, the fainter end of
the luminosity function should be determined in each case.

\acknowledgments

We would like to thank Ming Sun for making his metallicity data for 3C442A
available to us ahead of publication.  Also, we have benefitted from the
comments and conversations with Eric Bell, Jay Lockman, Mario Mateo, Doug
Richstone, Jon Miller, Trevor Ponman, Jesper Rasmussen, Steve Snowden, Richard
Mushotzky, Jimmy Irwin, and Renato Dupke. 
We gratefully acknowledge financial support from NASA under grants from the Chandra
and XMM programs, and from the NSF, as MEA is supported on a Graduate Fellowship.  This work
would not have been possible without the NASA Data Centers, most
notably HEASARC, ISRA, NED, along with the Chandra and XMM archives.

\clearpage



\begin{figure}
\epsscale{1.6}
\plottwo{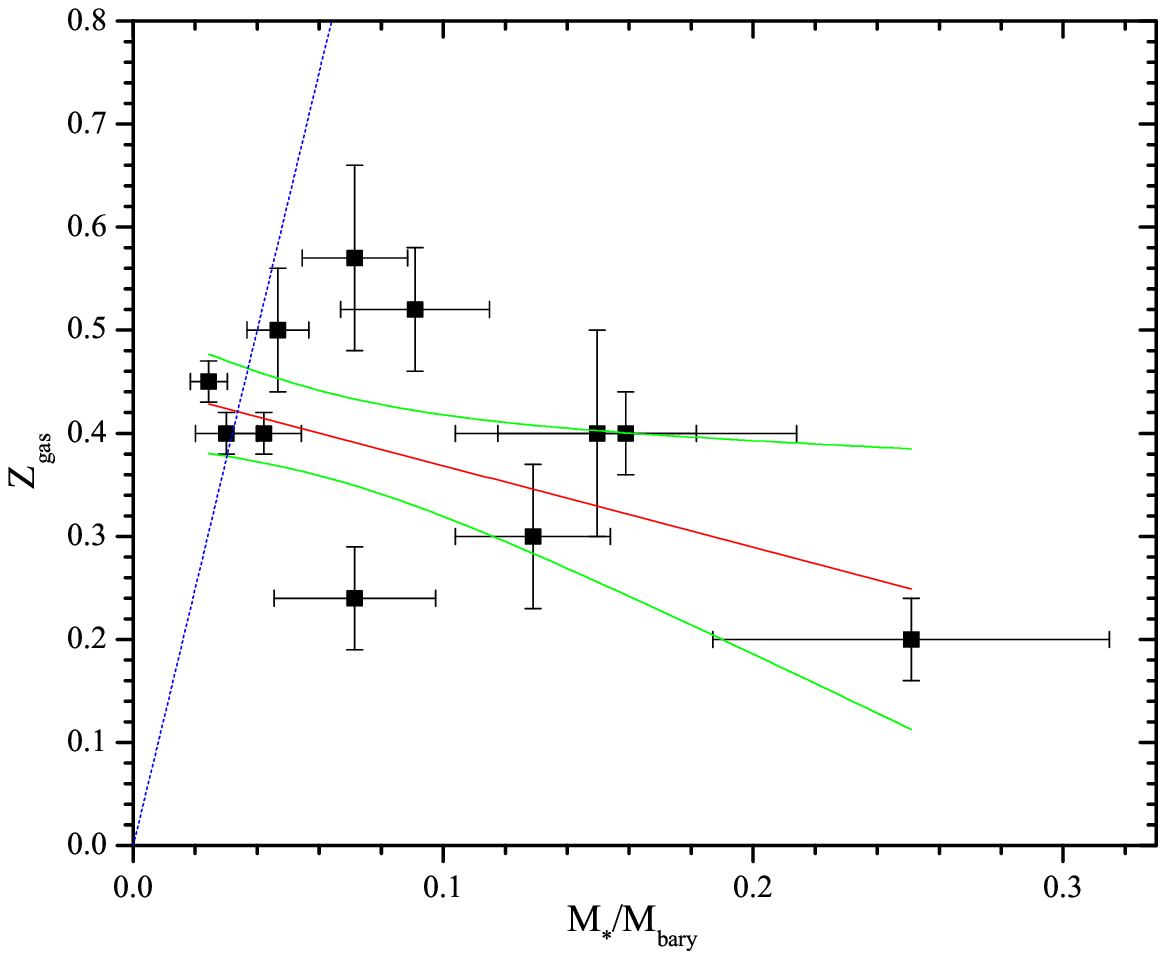}{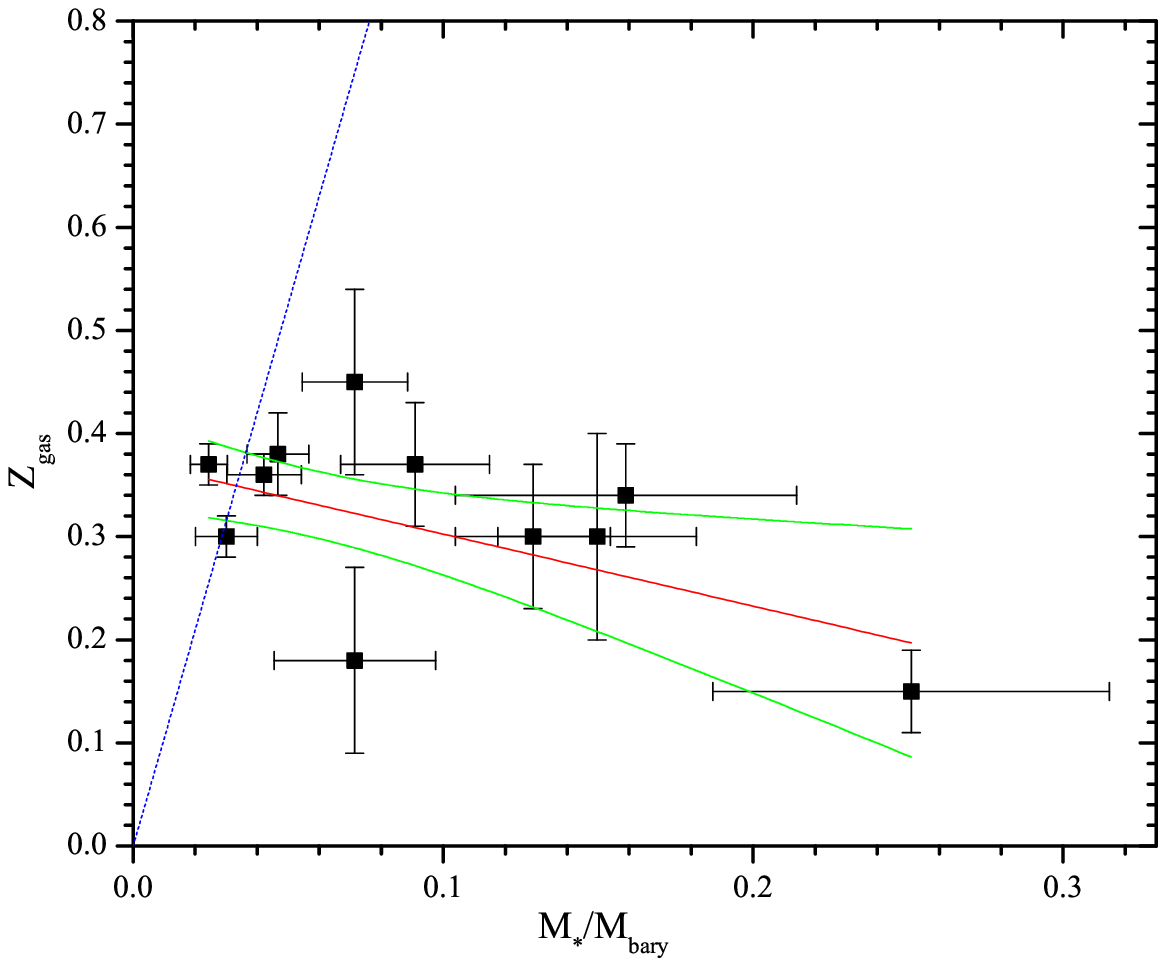}
\caption{Metallicity in the gas as a function of the stellar fraction of baryons, M$_*$/M$_{bary}$,
for a sample of 11 clusters with similar total baryon fractions, measured
relative to the cosmological value.  For the top panel, the metallicity is measured
for the whole cluster while for the lower panel, the metallicity is measured with
the cooling core excluded.  The steep dotted line, which is ruled out, is
the model predicted if the metallicity is proportional to the galaxies and passes through the rich clusters (the four clusters on the left).  
The red line is the best-fit linear relationship and the two curved green lines
are the 95\% confidence bounds.  The results imply that very little of the
metals in rich clusters (low M$_*$/M$_{bary}$) is due to galaxies.}
\end{figure}

\begin{figure}
\epsscale{1.}
\plotone{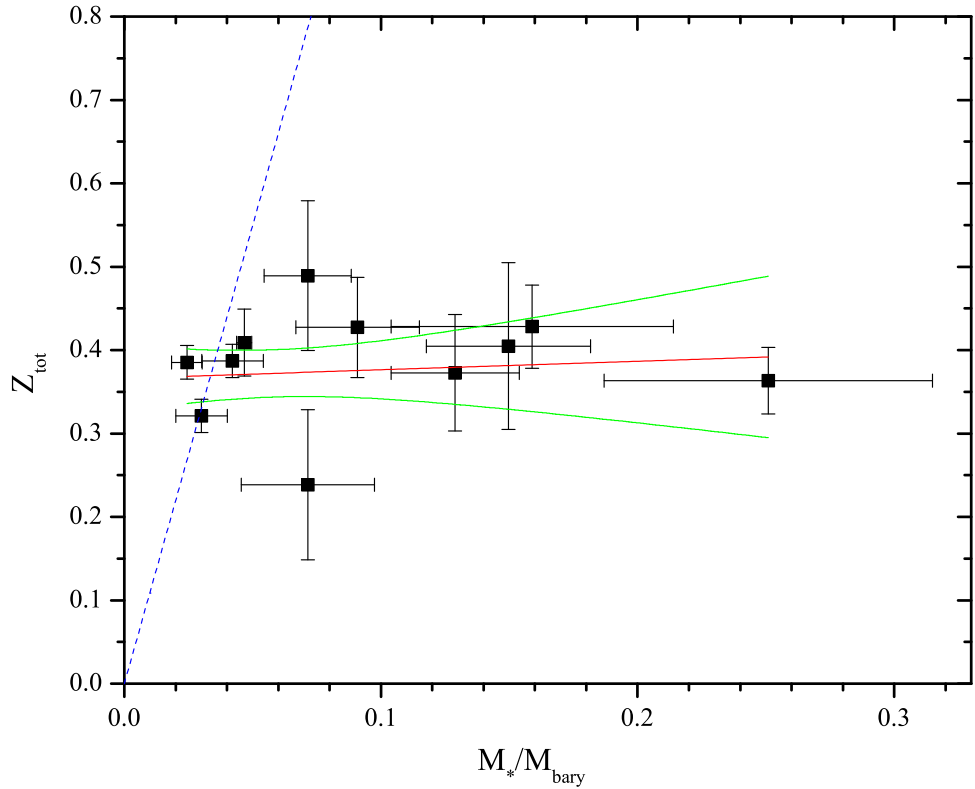}
\caption{This is the same as the above, except the metallicity of the stars is 
added as well (solar metallicity assumed for the stars).  The metallicity contribution 
from the gas is with the cooling core excluded.  This data set also rules out a 
model where the metallicity is proportional to the stellar fraction of baryons, M$_*$/M$_{bary}$.
}
\end{figure}

\clearpage

\begin{deluxetable}{llrrrrrrrrrrrr}
\tabletypesize{\scriptsize}
\rotate
\tablecaption{Galaxy Cluster Mass and Metallicity Contents}
\tablewidth{0pt}
\tablehead{
\colhead{Cluster} & \colhead{Alt Name} & \colhead{Redshift} & \colhead{T} & \colhead{M$_*$} & 
\colhead{M$_{tot}$(r$_{500}$)} & \colhead{b$_{frac}$} & \colhead{${\sigma}$} & \colhead{M$_*$/M$_{bary}$} & \colhead{${\sigma}$} & 
\colhead{Z$_{tot}$} & \colhead{${\sigma}$} & \colhead{Z$_{nc}$} & \colhead{${\sigma}$} \\
\colhead{} & \colhead{} & \colhead{} & \colhead{keV} & \colhead{M$_{\odot}$}  & \colhead{M$_{\odot}$}  &       & \colhead{} & \colhead{} & \colhead{} & \colhead{} & \colhead{} & \colhead{} &  }
\startdata
    Abell 133 &       & 0.0566 & 4.14  & 6.03E+12 & 3.13E+14 & 0.112 & 0.007 & 0.072 & 0.017 & 0.57  & 0.09  & 0.45  & 0.09 \\
    Abell 478 &       & 0.0881 & 7.94  & 1.20E+13 & 7.57E+14 & 0.156 & 0.014 & 0.042 & 0.012 & 0.40  & 0.02  & 0.36  & 0.02 \\
    Abell 1795 &       & 0.0625 & 6.12  & 4.57E+12 & 5.95E+14 & 0.132 & 0.007 & 0.024 & 0.006 & 0.45  & 0.02  & 0.37  & 0.02 \\
    Abell 1991 &       & 0.0587 & 2.61  & 3.72E+12 & 1.21E+14 & 0.141 & 0.010 & 0.091 & 0.024 & 0.52  & 0.06  & 0.37  & 0.06 \\
    3C442A &       & 0.0263 & 1.61  & 1.19E+12 & 3.90E+13 & 0.100 & 0.006 & 0.129 & 0.025 & 0.30  & 0.07  & 0.28  & 0.07 \\
    NGC 5098 & RGH 80 & 0.0362 & 1.05  & 2.20E+12 & 2.00E+13 & 0.190 & 0.021 & 0.251 & 0.064 & 0.20  & 0.04  & 0.15  & 0.04 \\
    Abell 160 &       & 0.0447 & 1.99  & 3.45E+12 & 7.40E+13 & 0.132 & 0.010 & 0.150 & 0.032 & 0.40  & 0.10  & 0.30  & 0.10 \\
    Abell S1101 & Sersic 159/03 & 0.058 & 2.69  & 1.51E+12 & 1.41E+14 & 0.147 & 0.025 & 0.030 & 0.010 & 0.40  & 0.02  & 0.30  & 0.02 \\
    Abell 1275 &       & 0.0603 & 1.63  & 3.72E+12 & 6.90E+13 & 0.144 & 0.030 & 0.159 & 0.055 & 0.40  & 0.04  & 0.34  & 0.05 \\
    Abell 2462 &       & 0.0733 & 2.62  & 2.04E+12 & 8.80E+13 & 0.135 & 0.012 & 0.072 & 0.026 & 0.24  & 0.05  & 0.18  & 0.05 \\
    Abell 2029 &       & 0.0773 & 8.47  & 1.15E+13 & 8.64E+14 & 0.157 & 0.011 & 0.047 & 0.003 & 0.50  & 0.06  & 0.38  & 0.05 \\
\enddata


\tablecomments{Only M$_{tot}$(r$_{500})$ is listed at r$_{500}$.  All other quantities are corrected to r$_{200}$. 
M$_*$ is the stellar mass, M$_{tot}$ is the total gravitating mass, b$_{frac}$ is the baryon fraction, 
M$_*$/M$_{bary}$ is the stellar to total baryon ratio, Z$_{tot}$ is the emission-weighted metallicity for the whole
cluster, and Z$_{nc}$ is the metallicity without the central cooling region.
}

\end{deluxetable}

\end{document}